# Current-Driven Ion-Cyclotron Waves in Presence of a Transverse DC Electric Fields in Magnetized Plasma with Charge Fluctuation

**Suresh C. Sharma**[*] **and Satoshi Hamaguchi**

Science and Technology Center for Atoms, Molecules, and Ions Control, Graduate School of Engineering, Osaka University, 2-1 Yamada-oka, Suita, Osaka 565-0871, Japan

*Currently on Leave from the Physics Department, GPMCE (G.G.S. Indraprastha University,
  Delhi),  INDIA

## ABSTRACT

The current-driven electrostatic ion-cyclotron (EIC) instability is studied in presence of a transverse dc electric fields in a collisional magnetized dusty plasma. We derive the appropriate charging equation self consistently by using the Vlasov equation.

## 1. INTRODUCTION

Fluctuations of the dust grain charge are found to be a source of wave damping or growth[1-2]. Current-driven EIC waves in presence of a transverse dc electric field in magnetized plasma without dust has been studied experimentally[3] and theoretically[4] earlier.

## 2. INSTABILITY ANALYSIS

Dusty plasma densities $n_{e0}$, $n_{i0}$, and $n_{d0}$, static magnetic field B is in the *z*-direction, charge, mass and temperature for three species are (-$e,m,T_e$), ($e,m_i,T_i$), and (-$Q_{d0},m_d,T_d$), collisional frequency (= $\nu_e$ for electrons and =0 for ions), dc electric fields $E_0$ is in the x-direction, drift of magnitude $v_E=cE_0/B$ is in the y-direction. In addition drift $v_{de}$ along the magnetic field direction.

**Basic Equations**

$$m_\alpha \frac{d\vec{v}_\alpha}{dt} = q_\alpha \vec{E} - \frac{e}{c}\vec{v}_\alpha \times \vec{B} - \frac{\nabla(p_\alpha)}{n_\alpha} - m_\alpha \nu_e \vec{v}_\alpha. \tag{1}$$

$$\frac{\partial n_\alpha}{\partial t}+\nabla.(n_\alpha \vec{v}_\alpha)=0 \qquad (2)$$

$$\frac{\partial f_\alpha}{\partial t}+\vec{v}.\nabla f_\alpha+\frac{1}{m_\alpha}[F_{ext}+q_\alpha(\vec{E}+\vec{v}_\alpha\times\vec{B})].\nabla f_\alpha=0 \qquad (3)$$

$$\nabla.\vec{E}=4\pi\rho_\alpha \qquad (4)$$

where α=e, i

**Charging equation[5]**

$$\frac{dQ_{d_1}}{dt}+\eta Q_{d_1}=-|I_{0e}|\frac{n_{e1}}{n_{e0}}+I_{i1}, \qquad (5)$$

By solving Eqs. (1)-(5), we obtain

$$\epsilon r(\omega,k) + i\, \epsilon i(\omega,k) = 0, \qquad (6)$$

where

$$\varepsilon_r(\omega,k)=1+\frac{\omega_p^2}{k_z^2 v_{te}^2}-\frac{\omega_{pi}^2}{(\omega_2^2-\omega_{ci}^2)}\frac{k_\perp^2}{k_z^2}+\frac{\beta\eta}{(\omega^2+\eta^2)}\frac{\omega_{pi}^2}{k_z^2 v_{te}^2}-$$
$$\frac{\beta\eta}{(\omega^2+\eta^2)}\frac{\omega_p^2 v_e \omega_1}{k_z^4 v_{te}^4}+\frac{\beta\eta}{(\omega^2+\eta^2)}\frac{\omega_{pi}^2}{(\omega_2^2-\omega_{ci}^2)}\frac{4a^2 n_{d0}}{k_z}\times$$
$$\left[-3+\frac{4e(\Phi_g-\Phi_0)}{m_i v_{ti}^2}\right]\frac{k_\perp^2}{k_z^2}, \qquad (7)$$

$$\varepsilon_i(\omega,k) = \frac{\omega_p^2 v_e \omega_1}{k_z^4 v_{te}^4} + \frac{\beta\omega}{(\omega^2+\eta^2)} \frac{\omega_p^2}{k_z^2 v_{te}^2} + \frac{\beta\eta}{(\omega^2+\eta^2)} \frac{\omega_p^2 v_e \omega_1}{k_z^4 v_{te}^4}$$
$$+ \frac{\beta\eta}{(\omega^2+\eta^2)} \frac{\omega_p^2}{(\omega_2^2-\omega_{ci}^2)} \frac{4a^2 n_{d0}}{k_z}[-3 + \frac{4e(\Phi_g - \Phi_0)}{m_i v_{ti}^2}] \frac{k_\perp^2}{k_z^2},$$

(8)

We write $\omega = \omega_r + i\gamma$ ( $|\gamma| \ll \omega_r$ ). The real and imaginary part of the frequencies are given by

$$\omega_r = k_y v_E + [\omega_{ci}^2 + k_\perp^2 c_s^2 \frac{n_{i0}}{n_{e0}}$$
$$- \frac{\beta\eta}{(\omega^2+\eta^2)} 4a^2 n_{d0} \frac{k_\perp^2}{k_z} c_s^2 \frac{n_{i0}}{n_{e0}}[-3 + \frac{4e(\Phi_g - \Phi 0)}{m_i v_{ti}^2}]^{1/2},$$

(9)

and

$$\gamma = -\frac{1}{2} \frac{n_{e0}}{n_{i0}} \frac{m_i}{m} [\{1 - \frac{k_z v_{de}(1+R)}{\omega_r}\} \frac{v_e \omega_r}{k_z^4 v_{te}^4}(1 + \frac{\beta\eta}{(\omega^2+\eta^2)}) + \frac{\beta\omega}{(\omega^2+\eta^2)} \frac{1}{(\omega_{2r}^2 - \omega_{ci}^2)}] \Bigg/$$

$$\frac{\omega_r(1-\frac{k_y v_E}{\omega_r})}{[(\omega_r - k_y v_E)^2 - \omega_{ci}^2]^2} \frac{k_\perp^2}{k_z^2} - \frac{\beta\eta}{(\omega^2+\eta^2)} \frac{n_{e0}}{n_{i0}} \frac{m_i}{m} \frac{v_e}{k_z^4 v_{te}^4} - \frac{\omega_r(1-\frac{k_y v_E}{\omega_r})}{[(\omega_r - k_y v_E)^2 - \omega_{ci}^2]^2},$$

(10)

$R = k_y v_E / k_z v_{de}$. If $n_{i0}/n_{e0} = 1$ and $n_{d0} \to 0$, i.e., $\beta \to 0$, we recover the expression for the growth rate and the dispersion relation [cf. Eqs. (23) and (24)] for the ion-cyclotron waves in presence of a transverse dc electric fields of Sharma et al.[4]

## 3. RESULTS

1. Dusty plasma parameters[6]

potassium ion plasma density $n_{i0} \approx 10^9$–$10^{10}$ cm$^{-3}$, $n_{e0} \approx 10^9$ cm$^{-3}$, relative density of negatively charged dust grains $\delta(=n_{i0}/n_{e0}) = 0$ to 8.0, $T_e \approx T_i \approx 0.2$ eV, $B = 0.1 \times 10^3$ G, $\omega_p = 1.78 \times 10^9$ rad/s, $\omega_{ci} = 0.23 \times 10^5$ rad/s (potassium), $n_{d0} \sim 5 \times 10^4$ cm$^{-3}$, $a \approx 1 \times 10^{-4}$ cm, $v_E = 3.3 \times 10^4$ cm/sec, $m_i /m$ (potassium) $\sim 1 \times 10^5$, $e(\Phi_g - \Phi_0)/m_i v_{ti}^2 = -1.71$, $k_y = 0.1$ cm$^{-1}$, $k_z = 1.0$ cm$^{-1}$ (i.e., $k_\perp^2 \ll k_z^2$). We have plotted in Fig. 1 the normalized real frequency ($\omega_r / \omega_{ci}$) of the Current driven EIC waves as a function of $\delta$. The normalized wave frequency ($\omega_r / \omega_{ci}$) increases about 10 % when $\delta$ changes from 1 to 4.0. Barkan et al.[6] have found that the wave frequency was some 10- 20% larger than the ion-cyclotron frequency in presence of dust. Hence our unstable wave frequency results, qualitatively and quantitatively, are similar to the experimental observations of Barkan et al. We have plotted in Fig. 2 the normalized growth rate ($\gamma / \omega_{ci}$) of the current driven EIC wave instability as a function of relative density of negatively charged dust $\delta$ for the same parameters as Fig. 1 and for $\nu_e \sim 1 \times 10^8$ sec$^{-1}$. The normalized growth rate ($\gamma / \omega_{ci}$) decreases monotonically with the relative density of negatively charged dust $\delta$.

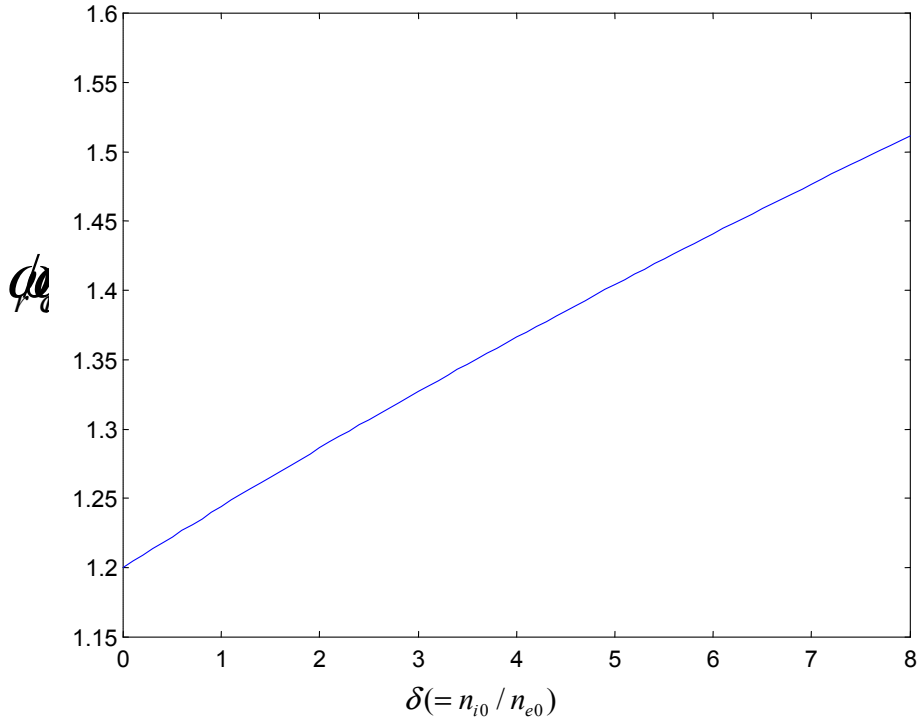

$\delta (= n_{i0} / n_{e0})$

Fig. 1 The normalized real frequency ($\omega_r / \omega_{ci}$) of the current driven ion-cyclotron waves as a function of the relative density of negatively charged dust grains δ.

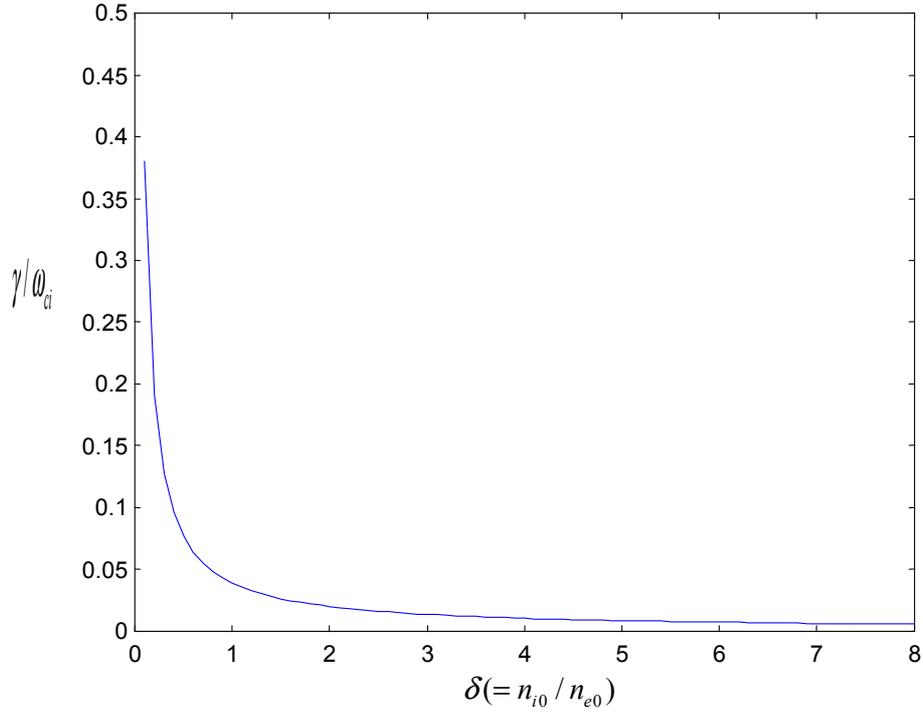

$\delta (= n_{i0} / n_{e0})$

Fig. 2 The normalized growth rate ($\gamma/\omega_{ci}$) of the current driven ion-cyclotron wave instability as a function of the relative density of negatively charged dust δ for the same parameters as Fig. 1 and for $\nu_e \sim 1\times 10^8$ sec$^{-1}$.